\documentclass[11pt]{article}
\usepackage{url}
\usepackage{footmisc}

\usepackage{amssymb}
\usepackage{moreverb}
\usepackage{caption}
\usepackage[a4paper, total={6in, 10in}]{geometry}

\newcommand\BibTeX{{\rmfamily B\kern-.05em \textsc{i\kern-.025em b}\kern-.08em
T\kern-.1667em\lower.7ex\hbox{E}\kern-.125emX}}
\usepackage{multirow}

\usepackage{graphicx}

\usepackage[T1]{fontenc}
\usepackage[utf8]{inputenc}

\usepackage{kpfonts}
\usepackage{lipsum}
\usepackage{authblk}

\title{DDoS Attacks: Tools, Mitigation Approaches, and Probable Impact on Private Cloud Environment}

\date{}
\author[1]{R.~K.~Deka\thanks{rup.deka@gmail.com}}
\author[1]{D.~K.~Bhattacharyya\thanks{dkb@tezu.ernet.in}}

\author[2]{J.~Kalita\thanks{jkalita@uccs.edu }}
\affil[1]{Department of Computer Science and Engineering, Tezpur University, Napaam, Assam, India,}
\affil[2]{Department of Computer Science, College of Engineering and Applied Science, University of Colorado, Boulder, CO, United States}

\begin{document}

\maketitle

\paragraph*{Abstract:}
The future of the Internet is predicted to be on the cloud, resulting in more complex and more intensive computing, but possibly also a more insecure digital world. The presence of a large amount of resources organized densely is a key factor in attracting DDoS attacks. Such attacks are arguably more dangerous in private individual clouds with limited resources. This paper discusses several prominent approaches introduced to counter DDoS attacks in private clouds. We also discuss issues and challenges to mitigate DDoS attacks in private clouds.


\section{Introduction and Related work}
The cloud computing infrastructure allows a service provider on the Internet to provide the use of computing resources to satisfy the demands of users. It also enables data centres and business organizations to provide benefits such as fast deployment, scalability, elasticity, security and resiliency. Virtualization makes it possible for cloud computing to provide services with optimal use of resources. Khorshed et al. \cite{khorshed2012survey} define cloud computing as ``a system of shared resources of a data center using virtualization technology. Such systems provide elastic on the basis of demand and ask for charges based on customer usage". 
\par
\begin{figure}
\captionsetup{font=normalsize}
\begin{center}
\includegraphics[scale=.7]{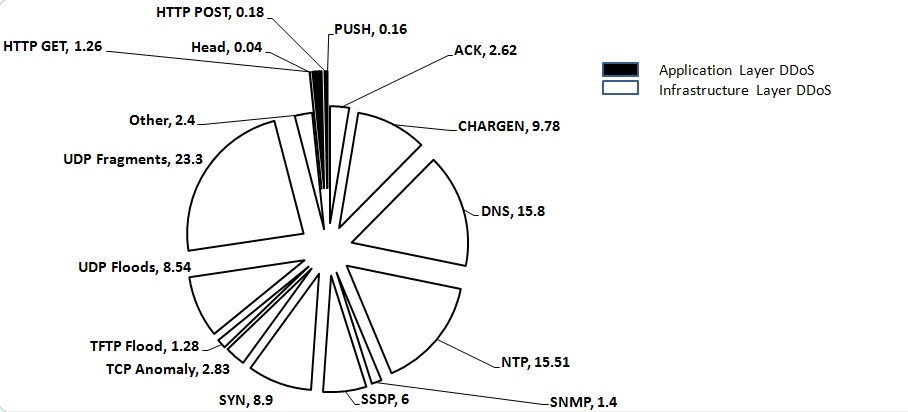} 
\caption{Intrusion Scenario in Quarter-2, 2016(in percentage)}
\label{fig:stat2} 
\end{center}
\end{figure}

\begin{figure}[!h]
\captionsetup{font=normalsize}
\centering

\includegraphics[scale=.5]{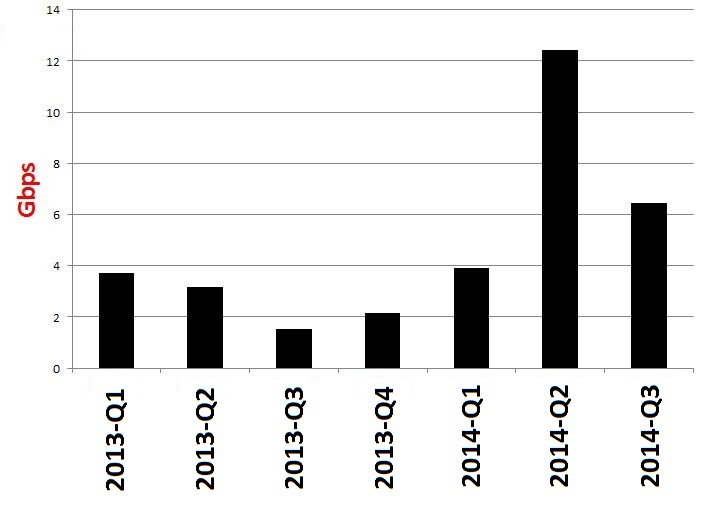}
\caption{Statistics on DDoS Attacks in 2016}
\label{fig:stat1} 

\end{figure}

 When providing relevant services on the Internet using a pool of shared resources, security is a major concern and policies must exist in cloud computing to address important issues such as reliability, security, anonymity and liability. Three types of intrusion can occur in a network of computing machines: scanning, DoS and penetration \cite{deka2015network}. The cloud incessantly faces security threats such as SQL injection, Cross Site Scripting (XSS), DoS (Denial of Service) and DDoS (Distributed Denial of Service) attacks, and hacking in general. The Arbor Networks\footnote{http://www.arbornetworks.com, Accessed: June, 2016} reported the largest (at that time) DDoS attack of 400 Gbps in 2014. In Figure \ref{fig:stat2}, DDoS attack trends in quarter two of year 2016 are shown\footnote{http://www.stateoftheinternet.com/securityreport, Accessed: November, 2016.}. In particular, large-scale DDoS attack frequency has continued to trend upward as shown in Figure \ref{fig:stat1}.

\par In October, 2016, the cyber attack that brought down much of America's Internet was caused by a new weapon called the Mirai botnet and was likely the largest of its kind in history. Unlike other botnets, which are typically made up of computers, the Mirai botnet is largely made up of so called Internet of Things (IoT) devices such as digital cameras and digital video recorder (DVR) players. The victims were the servers of Dyn, a company that controls much of the Internet's Domain Name System (DNS) infrastructure. It was hit on 21st of October, 2016 with an extraordinary attack strength of around 1.2 Tbps and remained under sustained assault for most of the day, bringing down many sites including Twitter, The Guardian, Netflix, Reddit, CNN and many others in Europe and US \footnote{https://www.theguardian.com/technology/2016/oct/26/ddos-attack-dyn-mirai-botnet, Accessed: February, 2017}.

%

\par Research on DDoS attacks and defense in the cloud environment is still at an early stage. These days, researchers are very much concerned about services in the cloud and cloud security. Sabahi \cite{sabahi2011cloud}, Pitropakis et al. \cite{pitropakis2013s}, and Grover and Sharma \cite{grover2014cloud} discuss efforts to secure user data in the cloud. Rather than storing the information locally at the client's infrastructure, information is stored in the cloud provider's location. It is obvious that in such a situation, people are worried about the security of their data. Thus cloud organizations should provide adequate security for the customer and also for the safety of their own.
\par In the context of the cloud, requests for resources like virtual machines (VMs) can be made by any user through the Internet. As a result, a network of zombies can easily launch DDoS attacks. Modi et al. \cite{modi2013survey} provide a survey of different types of intrusions which can take place in the cloud environment. Khorshed et al. \cite{khorshed2012survey}, and Subashini and Kavitha \cite{subashini2011survey} focus on flaws, challenges, security concerns in different service layers.
\par In this article, we discuss the seriousness of the threats posed by DDoS attacks in the context of the cloud, particularly in the individual private cloud. We present a discussion of different approaches which are used to defend or mitigate DDoS attacks in a general network architecture, and also some approaches that consider cloud computing technology in particular. Unlike \cite{modi2013survey}, we highlight challenges and issues faced particularly by the private cloud environment when facing DDoS attacks in a general way. A generic framework is discussed to defend against DDoS attacks in an individual private cloud environment taking into account different challenges and issues.

\par The first reported occurrence of a DDoS attack was from 1999 against servers at the  University of Minnesota. In the early 2000s, many popular and major Websites like Yahoo!, Ebay, CNN and Amazon.com were assaulted by DDoS attacks \cite{lau2000distributed}. Their systems were down for hours and users were denied access to services \cite{peng2007survey}. These attacks were able to create disaster because of the use of botnets. Stone-Gross et al. \cite{stone2009your}, and Hoque et al. \cite{hoque2014network} provide a detailed investigation of botnets \cite{fabian2007my}, a network of compromised machines under the control of a master. Khorshed et al. \cite{khorshed2012survey} provide a survey of challenges related to the cloud and present a proactive approach towards detection of attacks in the cloud.

\par A large number of methods have been documented and categorized in \cite{bhattacharyya2013network} to detect DDoS attacks. These methods or approaches are supervised learning, unsupervised learning, probabilistic learning, soft computing, and knowledge-based.
\par It is important to note that usually only two types of attacks are mounted depending on the traffic rate, i.e., high-rate and low-rate. If a numbers of legitimate users access the Internet at a high rate, and a sophisticated attacker attempts to mimic legitimacy like a flash crowd at the same time, it is tough to discriminate between them. Yu  et al. \cite{yu2012discriminating} formulate a feasible theory for distinguishing between the two using the concept of flow correlation coefficient. Xiang et al. \cite{xiang2011low} show how a low-rate attacker can take advantage of flaws in network protocols and also present a detection method for such attacks.

There has been some work on mitigating or tolerating DDoS attacks in the cloud environment. With increased sophistication of attackers, protection of open systems is increasingly challenging. Nguyen and Sood \cite{nguyen2011comparison} opine that intrusion tolerance should be a part of overall in-depth security. They compare three types of intrusion-tolerant system architectures. Lua and Yow \cite{lua2011mitigating} propose a method in which an intelligent large swarm network is used to mitigate the attack. The swarm network constantly reconfigures itself through the use of a parallel optimization algorithm such as the Intelligent Water Drop mechanism \cite{shah2009intelligent}. Amazon has created a technique called cloudWatch\footnote{https://aws.amazon.com/cloudwatch/, Accessed: August, 2016.} to monitor resources and to mitigate the situation according to the attack. Yu et al. \cite{yu2014can} attempt to provide the theory of optimal resource allocation in a cloud platform when defending a DDoS attack. Wang et al. \cite{wang2012measurement} have also developed a theory on optimal resource allocation, which is adaptable to the cloud scenario.

\begin{table*}[t]
\captionsetup{font=normalsize}
\normalsize
\begin{center}
\caption{Comparison with Existing Survey Articles}
\label{tab:comparison}
\begin{tabular}{|p{49mm}| c | c | c | c | p{19mm} | }
\hline
Authors & Year  & Attacks  &  Defense  & Issues  & Recommen- dations \\
   &  &  Included &   Solutions &  \& Challenges  & \\
\hline
&&&&&\\
Subashini and Kavitha \cite{subashini2011survey} & 2010   & $\checkmark$ & $\times$ & $\checkmark$ &  $\times$ \\
&&&&&\\
Sabahi \cite{sabahi2011cloud} & 2011  & $\checkmark$ & $\times$ & $\checkmark$ &   $\times$\\
&&&&&\\
Bhadauria et al. \cite{bhadauria2011survey} & 2011   & $\checkmark$ & $\times$ & $\checkmark$ &   $\times$\\
&&&&&\\
Khorshed et al. \cite{khorshed2012survey} & 2012   & $\checkmark$ & $\times$ & $\checkmark$ &  $\times$ \\
&&&&&\\
Modi et al. \cite{modi2013survey} & 2013   & $\checkmark$ & $\times$ & $\times$	&  $\times$ \\
&&&&&\\
Zhang et al. \cite{zhang2013survey} & 2013   & $\checkmark$ & $\times$ & $\times$ & $\times$ \\
&&&&&\\

Grover and Sharma \cite{grover2014cloud} & 2014  & $\checkmark$ & $\times$ & $\checkmark$ &  $\times$ \\
&&&&&\\
Wong and Tan. \cite{wong2014survey} & 2014  & $\checkmark$ & $\checkmark$ & $\checkmark$ &  $\times$\\
&&&&&\\
Our survey &    & $\checkmark$ & $\checkmark$ & $\checkmark$ & $\checkmark$\\
&&&&&\\
\hline
\end{tabular}
\end{center}
\end{table*}

\par In Table \ref{tab:comparison},  a comparison is provided among few existing survey papers with our work. For comparison, we choose four parameters, inclusion of attacks, description of defense solutions, issues and challenges, and inclusion of recommendations in these papers. khorshed et al. \cite{khorshed2012survey}, Grover and Sharma \cite{grover2014cloud}, and Subashini and Kavitha \cite{subashini2011survey} discuss that despite a lot of talk about the cloud, customers were still reluctant to deploy their business in the cloud. Security and complications with data privacy and data protection continue to restrict the growth of the cloud market and these survey papers are more specific to the security issues that have been raised due to the nature of the service delivery system of a cloud environment. Sabahi \cite{sabahi2011cloud} also raise the same concern about the cloud environment. Comparison between the benefits and risks of cloud computing is necessary for a full evaluation of the viability of cloud computing. Some critical issues that clients need to consider arise as they contemplate moving to cloud computing. Sabahi summarize reliability, availability, and security issues faced by cloud computing, and proposed feasible and available solutions for some of them. In a cloud computing environment, the entire data is deployed over a set of networked resources, and such data can be accessed through virtual machines. Since these data centers may be anywhere in  the world beyond the immediate reach and control of end users, there are many types security and privacy challenges that need to be understood and taken care of. There is always a possibility of server breakdown that has been witnessed often in recent times. Such things are extensively surveyed by Bhadauria et al. \cite{bhadauria2011survey}, who elaborate and analyze the numerous unresolved issues threatening the adoption of cloud computing and the diffusion affect the various stake holders linked to it.

\par Modi et al. \cite{modi2013survey} discuss different intrusions that affect availability, confidentiality and integrity of cloud resources and services. A few existing proposals including Intrusion Detection Systems (IDS) and Intrusion Prevention Systems (IPS) on the cloud are briefed. Many,commercial cloud providing businesses have emerged in the past deacde, and each one provides its own cloud infrastructure, APIs and application description formats to access the cloud resources and also support for Service Level Agreements (SLAs). As a result, vendor lock-in has seriously restricted the flexibility of end users, who would like to deploy applications over different infrastructures in different geographic locations, or to migrate a service from one provider's cloud to another. To enable seamless sharing of resources from a pool of cloud providers, efforts have emerged recently in both the industry and academia to facilitate cloud interoperability, i.e., the ability for multiple cloud providers to work together. Zhang et al. \cite{zhang2013survey} discuss all this and conduct the survey on the state-of-the-art efforts, with a focus on cooperation among different IaaS (Infrastructure as a Service) cloud platforms. They investigate the existing studies on taxonomies and standardization of cloud cooperation. Another big issue is that DDoS attacks today have been amplified into terabit volume with broadband Internet access and with the use of more powerful botnets. As a result, common DDoS mitigation and protection solutions implemented in small and large organizations' networks and servers are no longer effective. Wong and Tan \cite{wong2014survey} provide an in-depth study on the current largest DNS reflection attack with more than 300 Gbps strength on Spamhaus.org. They review and analyze the currently most popular DDoS attack types launched by hacktivists. Effective cloud-based DDoS mitigation and protection techniques proposed by both academic researchers and large commercial cloud-based DDoS service providers are also discussed.

\par This paper presents an organized survey concerning security in the network infrastructure of cloud computing, specifically impact of DoS and DDoS attacks on the networking services of a cloud environment. It begins with a description of types of cloud environments and then different types of DDoS attacks. It also highlights the seriousness of DDoS attacks in private clouds. We present an in-depth discussion of the challenges and issues in defending such attacks. The major contributions of this survey are the following.

%
%

\begin{figure*}[htbp]
\captionsetup{font=normalsize}

\includegraphics[scale=.65]{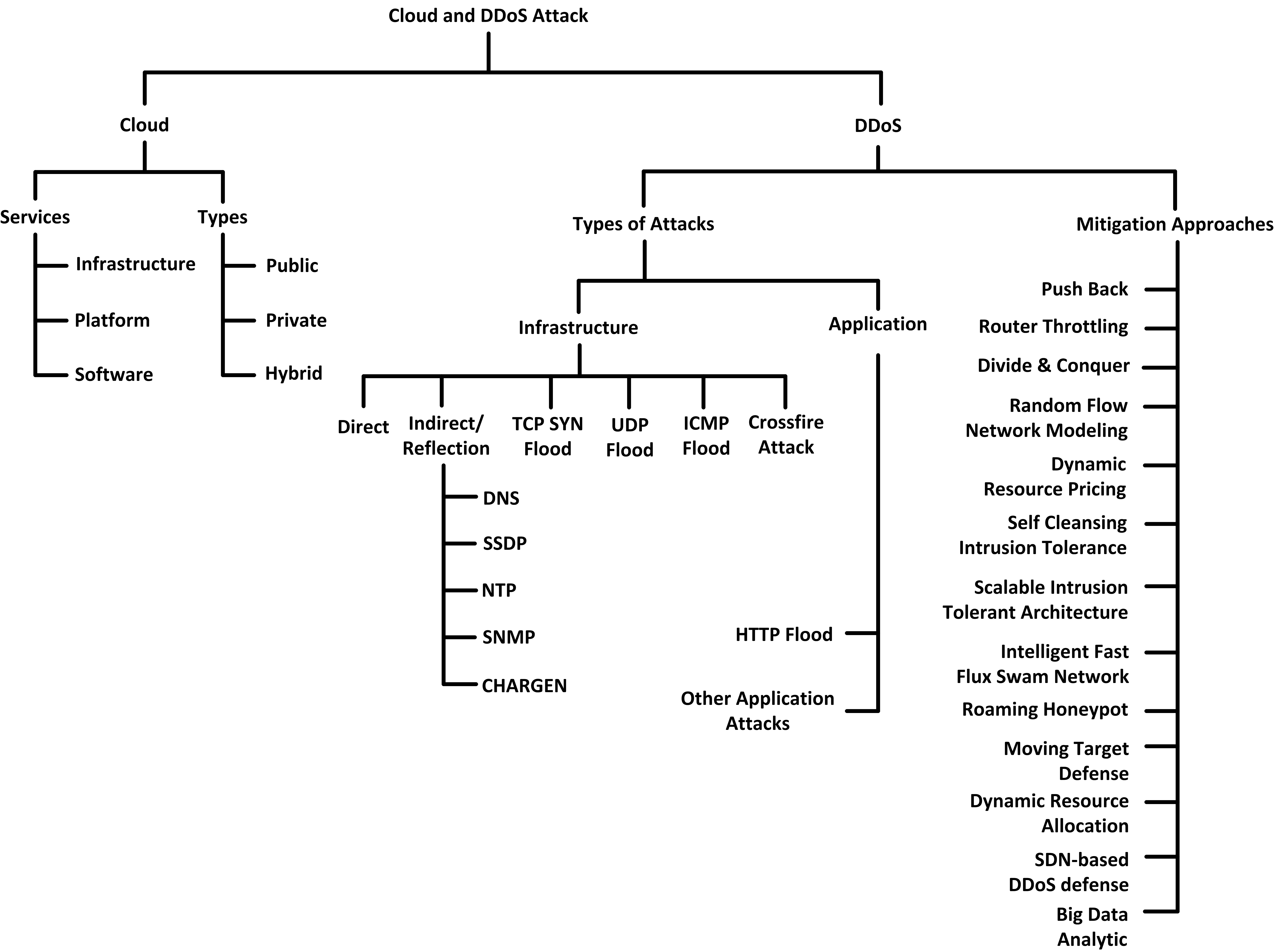}

\caption{A Taxonomy}
\label{fig:cloudtaxonomy} 
\end{figure*}

\begin{itemize}
\item Our presentation is specific to security of cloud computing. 
\item There are just a handful of surveys on cloud security, and published surveys do not emphasize the impact of DDoS attacks on individual private clouds. We present challenges and issues to help the researcher in creating a defense theory and in building a defense system against DDoS attacks.
\item Pros and cons analysis of a large number of detection and mitigation methods is included.
\item We also discuss trending concepts such as the role of big data and software defined networking in cloud security.

\item A generic framework for device defense mechanism in a cloud based environment is also presented.
\end{itemize}

\par The rest of the paper is organized as follows. Different deployment models of clouds, DDoS attacks, and types of DDoS attacks along with probable impact on private clouds are discussed in Section 2. Different existing approaches and potential solutions are briefed and some recommendations for developing a defense model are presented in Section 3. In Section 4, challenges and issues related to private cloud in defending against DDoS attacks are presented. A generic framework to defend against DDoS attacks is discussed in Section 5. Finally, we present conclusions in Section 6. In Figure \ref{fig:cloudtaxonomy}, a taxonomy of terms and concepts used in the entire article is provided for better understanding as the reader proceeds with the article.

\section{Cloud And DDoS Attack}
\label{sec:Deployment_Models}
\subsection{Cloud Deployment Models}

\begin{figure}[h]
\captionsetup{font=normalsize}
\centering
\includegraphics[scale=.45]{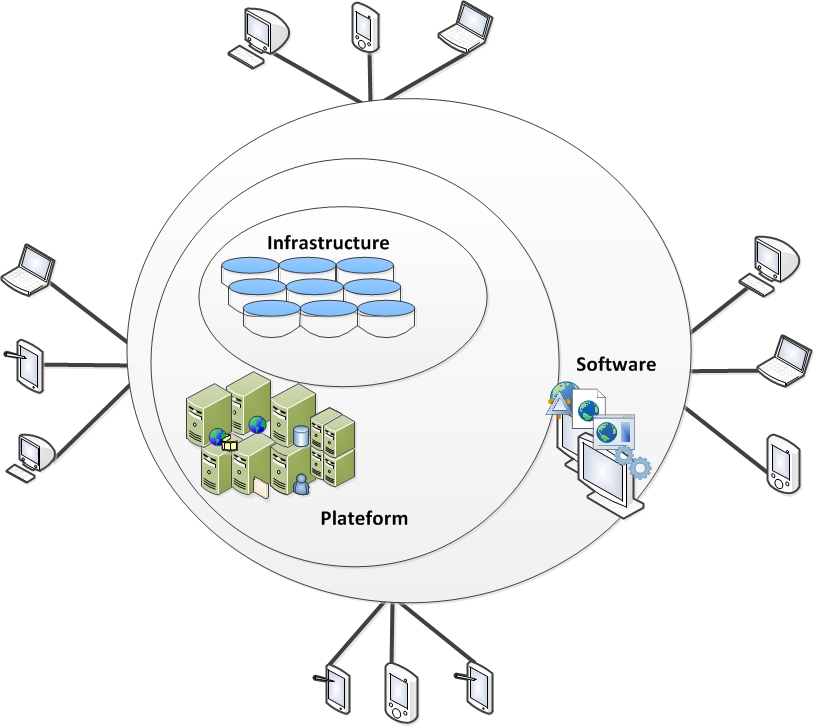}
\caption{A Cloud Node}
\label{fig:cloud} 
\end{figure}

A cloud node can provide three basic services to customers: infrastructure as a service (IaaS), platform as a service (PaaS) and software as a service (SaaS), as shown in Figure \ref{fig:cloud}. The deployment differences can be seen in the Figure \ref{fig:public_private}, and an explanation of different deployment models are given below.

\begin{figure}[h]
\captionsetup{font=normalsize}
\centering
\includegraphics[scale=.65]{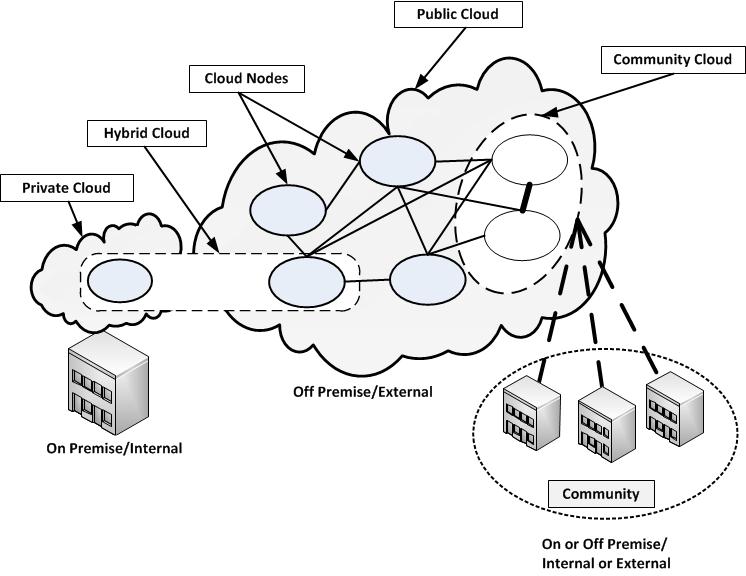}
\caption{Deployment Models}
\label{fig:public_private} 
\end{figure}

\begin{itemize}
\item[(a)] \textit{Public cloud}: The cloud is created for the general public where free or rental services are provided. This can be accessed by any authorized user. Examples of public clouds include Amazon Elastic Compute Cloud (EC2) \footnote{https://aws.amazon.com/ec2/, Accessed: July, 2016}, Google AppEngine\footnote{https://cloud.google.com/appengine/docs, Accessed: July, 2016} and Windows Azure Services Platform \footnote{https://azure.microsoft.com/en-in/, Accessed: August, 2016}. A public cloud provides abstractions for resources using virtualization techniques on a large scale. It benefits the user by providing for backup and access to secure resources by synchronizing, replicating and allocating the resources throughout the network.
\item[(b)]  \textit{Private cloud}: This type of cloud is managed internally by an organization or by a third party, and is hosted either internally or externally. A private cloud is designed to offer the same features and benefits of public cloud systems, usually with limited resources for maintaining the cloud environment. Unlike a public cloud, a private cloud remains within the corporate firewall, which means private cloud is privately manged by a company for private use of its private users and not for public on pay per use basis. Also, a private cloud can be used by a company to store sensitive data internally and at the same time provide the advantages of cloud computing within their business infrastructure, such as on demand resource allocation as in Apache CloudStack\footnote{https://cloudstack.apache.org/, Accessed: September, 2016}, OpenStack\footnote{https://www.openstack.org/, Accessed: September, 2016}, VMware vCloud Suite\footnote{http://www.vmware.com/in/products/vcloud-suite, Accessed: August, 2016}, etc. Individual private cloud customers as well as the provider (referred to as parties hosting their services in a cloud) do not have sufficient resources to handle a rapid increase in service demands. On the other hand, they too can avail the advantage of unique features of clouds.

\item[(c)]  \textit{Community Cloud}: This cloud infrastructure is provided for a specific or exclusive community of consumers\footnote{http://nvlpubs.nist.gov/nistpubs/Legacy/SP/nistspecialpublication800-145.pdf, Accessed: August, 2017}. These group of users or organisations may have shared concerns (e.g., mission, security requirements, policy, and compliance considerations). Community Cloud can be controlled by one or more organizations of that community, or a third party, or some combination of them. Also it can be exist on or off premises. some examples of community clouds are Dimension Data\footnote{http://www2.dimensiondata.com/services/cloud-services/provider-and-community-cloud, Accessed: August, 2017}, LayerStack\footnote{https://www.layerstack.com/cloud-servers}, and Zoolz\footnote{http://www.zoolz.com/overview/}.

\item[(d)]  \textit{Hybrid cloud}:  This is a combination of two or more linked cloud deployment models with a provision to transfer data between them. The combination may include both private and public clouds. For example, a company can maximize its efficiency by deploying public cloud services for all non-sensitive operations, but only deploy private cloud when it needs to store sensitive operations as it is surrounded by firewall, and ensure that all of their platforms are seamlessly integrated. This type of mixed cloud environment adds complexity to the distribution of applications across environments. Amazon Web Services\footnote{https://aws.amazon.com, Accessed: October, 2016}, Rackspace Hybrid Cloud\footnote{https://www.rackspace.com/en-in/cloud/hybrid, Accessed: October, 2016}, EMC Hybrid Cloud\footnote{https://www.emc.com/en-us/cloud/hybrid-cloud-computing/index.htm, Accessed: July, 2016}, HP Hybrid Cloud\footnote{http://www8.hp.com/in/en/cloud/helion-hybrid.html, Accessed: September, 2016} are some examples of hybrid clouds.
\end{itemize}

\begin{table}[!h]
\normalsize
\captionsetup{font=normalsize}
\begin{center}
\caption{Differences between private and public cloud}
\label{tab:diff_private_public}
\begin{tabular}{| p{2.5cm} | c | c |}
\hline
Key Points & Private & Public\\
\hline
Use of Technology & Old & New \\
Capital Expenses & Not Shifted & Shifted to \\
& & Operational \\
& & expenses\\
Utilization Rate & Low & High \\
Infrastructure Cost & High & Low \\
Elasticity & Less & More \\
Economies of Sale & Less & High \\
Business Attraction & Low & High \\
Security & Less & High \\
Perimeter Complacency & Suffer & Not suffer\\
Skill Level & Unknown & Usually High \\
Penetration Testing & Insufficient & Sufficient \\
Business Focus & Deeply in  & Out of \\
& Data center &  Data center\\
\hline
\end{tabular}
\end{center}
\end{table}

\subsubsection{Differences Between Private Cloud and Public Cloud}
In Table \ref{tab:diff_private_public}, differences between private and public clouds are enumerated. A cloud has shared general features, whether private or public. As clouds have evolved on and from the Internet, we can build defense models based on research that has been conducted on general defense solutions against DDoS attacks and features of clouds. We can then proceed to discuss individual private cloud defense. Private clouds require more attention because they have limited resources and the cost is high during an attack compared to a public cloud. This is because we know that a private cloud is accessed by authorized users or private organizations paying money as per need. Both ends (customer and service provider) heavily rely on security. A DDoS attack can cripple the whole private cloud and jeopardize whole businesses. So DDoS attack is more threatening to individual private cloud customers than a public cloud's customers.

\subsection{DDoS Attacks}
\label{sec:DDoS_Attacks_Cloud}

DoS attacks are intended to deny legitimate users access to network resources. Attackers create a botnet of compromised nodes on the Internet to support a DoS attack to inflict severe damage to a target. Such coordinated and distributed attacks are termed DDoS attacks. It is obvious that in the cloud environment, there are a lot of concentrated resources and the infrastructure is shared by a large number of users. A DDoS attack has the potential to do immense harm, much more than the harm that can be wrought upon single tenanted architectures \cite{DNet2013}. 

\begin{figure*}[h]
\captionsetup{font=normalsize}
\centering
\includegraphics[scale=.55]{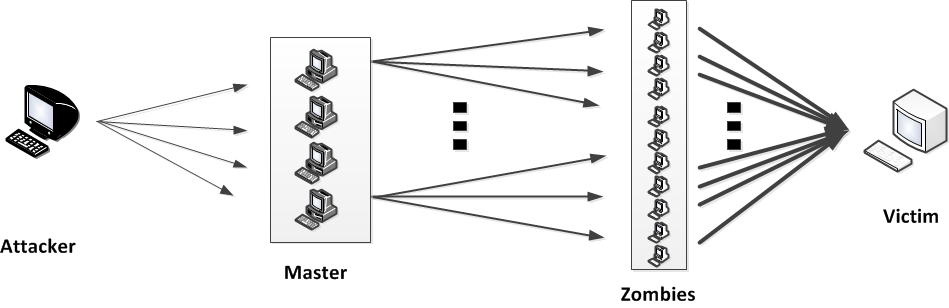}
\caption{Direct DDoS Attack}
\label{fig:sub1}
\end{figure*}%

\begin{figure*}[h]
\captionsetup{font=normalsize}
\centering
\includegraphics[scale=.55]{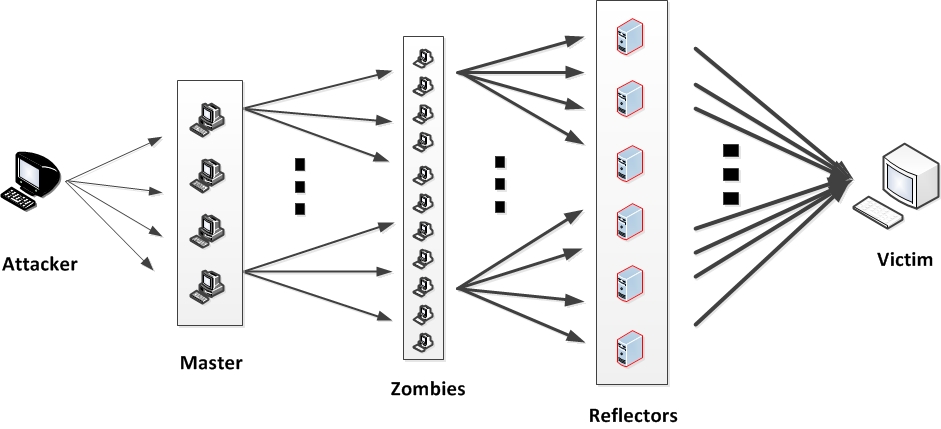}
\caption{Reflection/Indirect DDoS Attack}
\label{fig:sub2}
\end{figure*}

\subsubsection{Infrastructure level attacks}
Network bandwidth, routing equipment and computing resources are considered infrastructure. In this attack, the intruder attempts to overwhelm the resource capacity of a private cloud's infrastructure by sending a large number of fake requests, which exploit the limitation of a specific application to cause performance degradation or ultimately crash remote servers. Some commonly used infrastructure level attacks are listed below.

\par {\it a) Direct:} A direct Denial-of-Service attack is characterized by an explicit attempt to prevent the legitimate use of a service  \cite{mirkovic2004taxonomy}. A Distributed Denial-of-Service attack deploys multiple attacking entities to attain this goal as shown in Figure \ref{fig:sub1}. A DDoS attack includes an overwhelming quantity of packets sent from multiple attack sites to a victim site. These packets arrive in such a high quantity that some key resource at the victim is quickly exhausted. The victim either crashes or spends so much time handling the attack traffic that it cannot attend to its real work.

\par {\it b) Reflection/Indirect:} It is a type DoS attack in which multiple compromised victim machines unwillingly participate in a DDoS attack. Flashes of requests to the victim host machines are redirected or reflected from the victim hosts to the target. Some reflection or indirect based attacks are mentioned below. The general approach is as shown in Figure \ref{fig:sub2}.

\begin{itemize}
\item DNS (Domain Name Service) reflection or amplification attacks use botnets that send a large number of DNS queries to open DNS resolver using spoofed IP addresses of victims to produce an overwhelming amount of traffic with very little effort. Thus, such an attack can do a lot of damage as it is difficult to stop this type of attack at an early stage.
\item SSDP (Simple Service Discovery Protocol) reflection attacks are created using the Simple Object Access Protocol (SOAP) to deliver control messages to universal plug and play (UPnP) devices and to communicate information. These requests are created to elicit responses, which reflect and amplify a packet and redirect responses towards a target.
\item NTP (Network Time Protocol) reflection attacks are created by the attacker to send a crafted packet in which requests for a large amount of data are sent to the host. NTP is used to synchronize the time between client and server.
\item In an SNMP (Simple Network Management Protocol) reflection attack, the culprits send out a huge number of SNMP queries with forged IP addresses to numerous victim machines. SNMP is a network management protocol for configuring and collecting information from servers.
\item CHARGEN (Character Generator Protocol) is often misused when attackers use the testing features of the protocol to create malicious payloads and reflect them by spoofing the address of the source to direct them to the target. CHARGEN is a debugging and measurement tool and also a character generator service.
\end{itemize}
\par {\it c) TCP SYN flood:} Manipulating the 3-way handshake in a TCP connection, an attacker sends a lot of ordinary SYN segments to fill up resources causing a service to be denied for legitimate connections.
\par {\it d) UDP flood:} In this attack, massive amounts of
UDP packets are sent to random ports on the victim side. Sometimes ports remain open without knowledge of administrators, causing the server to respond. A response to each UDP packet with an IMCP unreachable reply to the spoofed source IP address makes the situation worse by overwhelming the network environment of the victimized IP addresses.
\par {\it e) ICMP flood:} 
ICMP flood, occasionally referred to also as a Smurf attack or Ping flood, is a ping-based DoS attack that sends large numbers of ICMP packets to a server and attempts to crash the TCP/IP stack on the server and cause it to stop responding to incoming TCP/IP requests.

\par {\it f) Crossfire Attack:} A botnet can launch an attack with low intensity traffic flows that cross a targeted link at roughly the same time and flood it. For example, a botnet controller can compute a large set of IP addresses whose advertised routes cross the same link, and then direct its bots to send low-intensity traffic towards these addresses. This type of attack is called the Crossfire attack \cite{kang2013crossfire}.

\subsubsection{Application level attacks}
Application layer DDoS attacks continue to grow in both complexity and prevalence. 

\begin{itemize}
\item[(i)]  \textit{Common application-layer DDoS attack types}: When a heavy amount of legitimate application-layer requests or normal requests that consume large amounts of server resources or high workload requests across many TCP sessions are sent to the server, they can cause common application layer DDoS attacks.
\item[(ii)]  \textit{HTTP flood attacks}: Some application level DDoS attacks come in the form of HTTP GET floods. HTTP request attacks are those attacks where attackers send HTTP GETs and POSTs to Web servers in an attempt to flood them by consuming a large amount of resources. The HTTP POST method enables attackers to POST large amounts of data to the application layer at the victim side, and it happens to be the second most popular approach among the application layer attacks. 
\end{itemize}

\subsection{Probable Impact of DoS/DDoS on Cloud}
As mentioned earlier, the cloud computing market continues to grow, and the cloud platform is becoming an attractive target for attackers to disrupt services and steal data, and to compromise resources to launch attacks. Miao et al. \cite{miao2015dark} present a large-scale characterization of inbound attacks towards the cloud and outbound attacks from the cloud using three months of NetFlow data in 2013 from a cloud provider. They investigate nine types (TCP SYN flood, UDP flood, ICMP flood, DNS reflection, Spam, Brute-force, SQL injection, Port scan, and Malicious Web activity (TDS)) of attacks ranging from network-level attacks such as DDoS to application-level attacks such as SQL injection and spam. Cloud computing features a cost-efficient, ``pay-as-you-go'' business model. A cloud platform can dynamically clone virtual machines very quickly, e.g., by duplicating a gigabyte level server within one minute \cite{peng2012vdn}. Despite the promising business model and hype surrounding cloud computing, security is the major concern for a business that is moving its applications to clouds. When a DDoS attack is launched from a botnet with a lot of zombies, Web servers can be flooded with packets quickly, and memory can be exhausted quickly in an individual private cloud. So, we can say that the main competition between DDoS attacks and defenses is for resources. The increase of DDoS attacks in volume, frequency, and complexity, combined with the constant required alertness for mitigating Web application threats, has caused many Website owners to turn to Cloud-based Security Providers (CBSPs) to protect their infrastructure  \cite{vissers2015maneuvering}. In one recent analysis\footnote{http:/www.cloudsecurityalliance.orgtopthreats, Accessed: November, 2016}, DDoS attacks are considered one of the top nine threats to cloud based environments. This report concludes that cloud services are very tempting to DDoS attackers who now focus mainly on private data centers. It is safe to assume that, as more cloud services come into use, DDoS attacks on them will become more commonplace. Some key findings are provided by InfoWorld\footnote{http://www.infoworld.com/d/cloud-computing/cloud-use-grows-so-will-rate-of-DDoS-attacks-211876, Accessed:October, 2016}, in 2013. 
\begin{itemize}
\item 94 percent of data center managers reported some type of security attacks.
\item 76 percent had to deal with distributed denial-of-service (DDoS) attacks on their customers.
\item 43 percent had partial or total infrastructure outages due to DDoS attacks.
\item 14 percent had to deal with attacks targeting a cloud service.
\end{itemize}
\par Unfortunately, the counterparts of clouds, e.g., client-server and peer-to-peer computing platforms, do not usually have sufficient resources to beat DDoS attacks. The public cloud infrastructure stands a better chance because a public cloud usually has a lot of resources that make it easy to handle a rapid increase in service demands to counter the attack dynamically. It is almost impossible to shut down such clouds by attacking them. But, if an intense DDoS attack occurs on customers of an individual private cloud like a data center with limited resources, it cannot escape from the DDoS attack, and it becomes a battle of survival using all the resources there are to confront. The essential requirement to defeat a DDoS attack is to allocate sufficient resources to mitigate attacks no matter how efficient our detection and filtering algorithms are.

\par Cloud Service Providers (CPS) usually provide cloud customers two resource provisioning plans: short-term on-demand and long-term reservation. Giant cloud providers, like Amazon EC2 and GoGrid, provide both plans \cite{chaisiri2012optimization}. If a customer chooses the first plan, it is charged based onresources used. This business model for resources is vulnerable to an Economic Denial of Sustainability (EDoS) attack \cite{idziorek2013insecurity,sqalli2011edos,somani2015ddos}. This kind of attack also disturbs the service of clouds that allocate resources based on spot instance \cite{wang2012cloud,yi2012monetary}. On the other hand, if a customer chooses the reservation plan, it makes a prior reservation for resources for the maximum usage for the business. In other words, the reserved resources for the application are limited from start. As a result, a threat of DDoS attack remains.

\begin{figure*}[ht!]
\captionsetup{font=normalsize}
\begin{center}
\includegraphics[scale=.6]{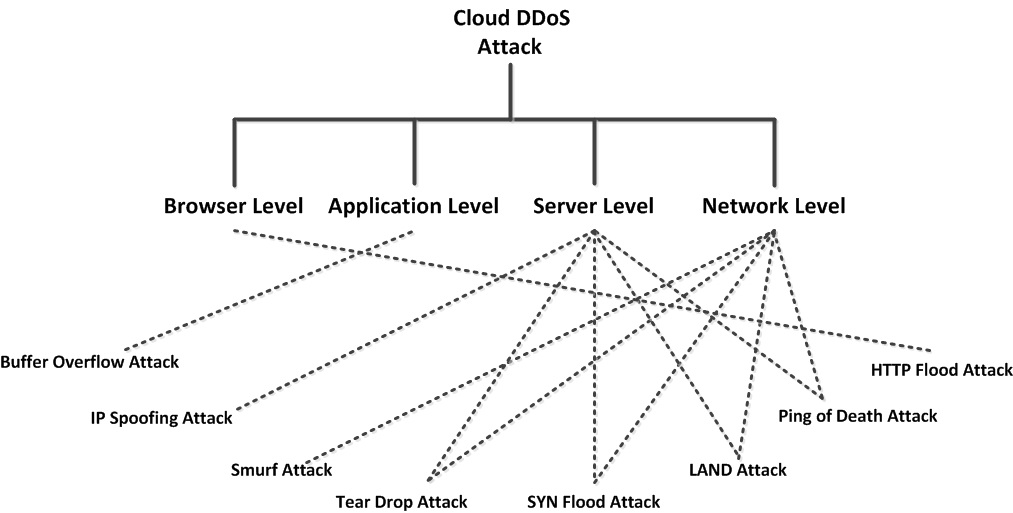}

\caption{Possible Scenario of DDoS Attack Types in Private Cloud}
\label{fig:attack_scenario} 
\end{center}
\end{figure*}

\par Some possible examples of DDoS attacks in cloud environments are Smurf attack, IP spoofing attack, Tear drop attack, SYN flood attack, ping of death attack, Buffer overflow attack, LAND attack, etc., as shown in Figure \ref{fig:attack_scenario} \cite{yan2015distributed,deshmukh2015understanding}. From many news report we can state that large-scale IoT-enabled DDOS attacks will continue to dominate enterprise security. Darwish et al.
\cite{darwish2013cloud} discuss DDoS attacks as attacks that target the resources of these services, lowering their ability to provide optimum usage of the network infrastructure. Due to the nature of cloud computing, the methodologies for preventing or stopping DDoS attacks are quite different compared to those used in traditional networks, and new approaches published till now are usually adapted versions of older approaches. In the above mentioned papers, we can find descriptions about the effect of DDoS attacks on cloud resources and recommend practical defense mechanisms against different types of DDoS attacks in the cloud environment.

\subsection{Discussion}
We summarize below the security concerns in the private cloud against DDoS attacks in the following.
\begin{itemize}
\item Since infrastructure is shared by a large number of clients, a massive DDoS attack potentially has great impact.

\item Symptoms of DoS or DDoS attacks are unusually slow network performance, unavailability of a particular Website, inability to access any Website, and dramatic increase in the amount of spam.

\item The patterns of DDoS attack are always changing. Attack growth, intensity and penetration time change fast along with the Internet world.

\item In the resource constrained environment of a private cloud network, it is essential to handle a DDoS attack as quickly as possible.

\item It is usually a battle for survival with all the resources the private cloud can muster.

\item Deft resource management is necessary to defend against a DDoS attack in the cloud when a DDoS attack is mounted against a private cloud, especially in an individual private cloud. Putting the best detection or filtering algorithm may not always work. But tolerating the attack by optimal resource utilization may resist the attack and may help counter the DDoS attack.

\item Virtualization of resources gives some edge over DDoS attacks in a cloud environment.

\end{itemize}

\section{Mitigation Approaches}
\label{sec:mitigation}
\begin{figure}[t]
\captionsetup{font=normalsize}
\begin{center}
\includegraphics[scale=.7]{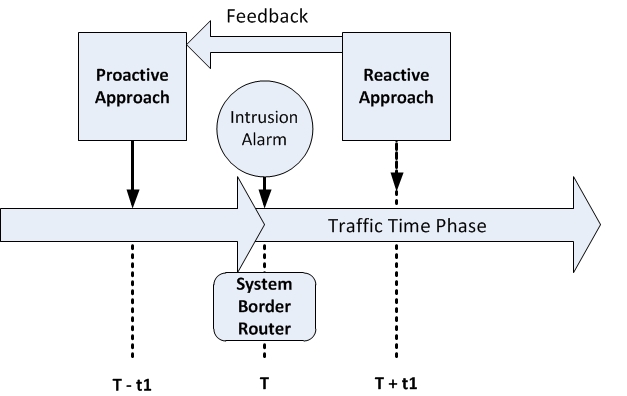}

\caption{Attack Response Scenario}
\label{fig:attack_response} 
\end{center}
\end{figure}

Mitigating DDoS attacks is a classic problem. However, in the cloud environment, it becomes a bigger challenge \cite{anwar2014can}. We also cannot totally separate a cloud environment from the traditional network infrastructure.  Though the data center networks are more complex in reality, the backbone infrastructure is based on the traditional network architecture. These complex networks are adapted for virtualization for scalability and robustness. We present several approaches that have been applied in network contexts in transition so that we can understand the requirements, issues and challenges in building defense modules against DDoS attacks in a data center-like private cloud environment. One can see the evolving nature of the defense approaches along with the evolution of the Internet in the discussed approaches. All approaches presented in this section have some advantages, which can be adapted for private cloud-like environment. Some promising new approaches have also been developed in the context of the cloud. These include like SDN-based ideas and ideas from the big data analytic point of view \cite{fayaz2015bohatei}. A defense approach can be deploy in the network itself or in the host(victim) environment. We analyze different existing approaches, and based on features of the approaches such as the level of operation, time to respond, and time to cooperate with other devices, we divide active response into two main categories, as shown in Figure \ref{fig:attack_response}.

\begin{itemize}

\item In a {\it proactive} approach, steps taken to control potential incident activity before it happens rather than waiting for it to happen.

\item A {\it reactive } approach detects the abnormality and informs the security administrator or automatically takes a responsive counter-action immediately i.e., in real-time. A reactive response reacts only after the intrusion is detected.
\end{itemize}

In the rest of this section, a few prominent approaches are discussed. The models developed by different authors based on these approaches are analyzed. Each of the methods can be included either in the proactive or the  reactive category. It very much remains open to debate which type category will work best in the individual private cloud environment.

\par{{\it\textbf{Push-back}}:}
Mitigating DDoS attacks is a congestion-control problem because most congestion happens due to malicious hosts not obeying traditional end-to-end congestion control policies. Most researchers think that the problem needs to be handled by routers. Functionality can be introduced in each router to detect and preferentially drop malicious packets, which probably belong to an attack. A push-back mechanism based on managing congestion at the routers has been implemented by Ioannidis and Bellovin \cite{ioannidis2002implementing}. Their architecture has three main parts. Congestion signature matching tries to monitor packets in the incoming queue. Matching patterns are then sent to a rate-limiter to decide whether a packet is to be dropped or forwarded. The packets to be dropped are sent to the \textit{Push-Back Daemon}, which periodically updates congestion signatures and the rate limit in the rate-limiter. A cooperative environment among neighboring routers implementing this approach will be appropriate for dealing with DDoS attacks in the cloud environment and provide for a dynamic solution in real time.

\par
\begin{figure}[h]
\captionsetup{font=normalsize}
\begin{center}
\includegraphics[scale=.7]{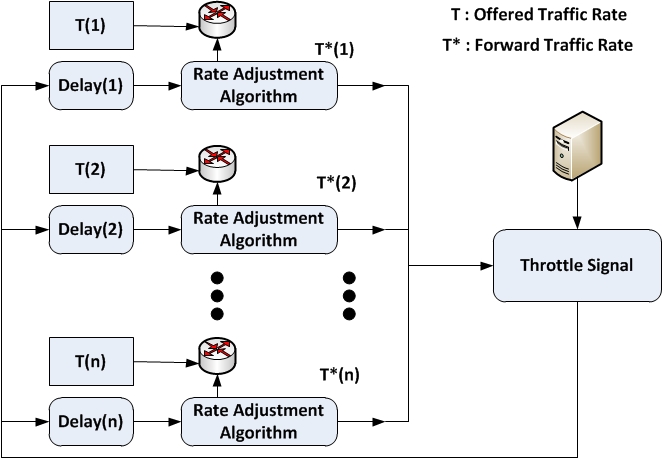}

\caption{Router Throttling Model Proposed by Yau et al.\cite{yau2005defending}}
\label{fig:throttle} 
\end{center}
\end{figure}

{{\it\textbf{Router Throttling}}:}
The basic concept behind router throttling is to develop a model to throttle or control the flow of traffic at upstream routers of a server, which may be under stress or attack. It is a proactive process to forestall an impending attack. Participating routers can regulate the packet rate destined for a server. Yau et al. \cite{yau2005defending} propose and simulate a router throttling model to establish the efficacy of the concept, as shown in Figure \ref{fig:throttle}. This proactive process may be very useful in the private cloud environment, because it can reduce computation load in an end server with limited resources. This idea can also increase the service reliability for legal users. Using the improvised K-level max-min fairness theory \cite{nace2008max}, Yau et al. find that the throttling mechanism is highly effective in countering an aggressive attacker. They efficiently regulate the server load to a level below its design limit in the midst of a DDoS attack.

\par{{\it\textbf{Divide and Conquer}}:}
Chen et al. \cite{chen2007divide} use the divide and conquer strategy to actively throttle attack traffic. They present a diagnosis and attack mitigation scheme that combines the concepts of push-back and packet marking. Attack detection is performed near the victim and packet filtering is executed close to attack sources. Initially, the intrusion detection system detects the attack on the victim side. The victim end instructs the upstream router to mark malicious packets with trace back information to filter out bad packets when they arrive again at the victim later. The traceback scheme is carried out till the source end is reached. We believe that this idea can be adapted to the cloud environment.

\par{{\it\textbf{Random Flow Network Modeling}}:}
This approach adapts the theoretical concept represented by the max-flow min-cut theorem of \cite{dantzig2003max} concerning flow in a network. Kong et al. \cite{kong2003random} rely on this theory in designing a random flow network model to mitigate DDoS attacks. They show that this mitigation problem can be reduced to an instance of the maximum flow problem. We know that a DDoS attacker heavily pumps the flow of traffic towards the sink. The strategy depends on the fact that the maximum achievable flow value from the source to the sink is equal to the capacity of a certain cut in the flow network. This method is suitable for any kind of computing environment because it does not depend on the end infrastructure, rather it is concerned with the intermediate network infrastructure.

\begin{figure}[h]
\captionsetup{font=normalsize}
\begin{center}
\includegraphics[scale=.65]{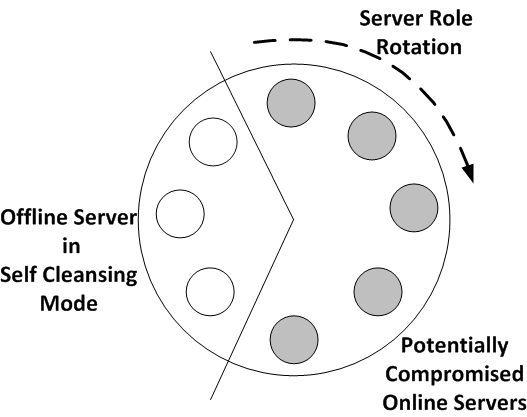}

\caption{A High-Level View of SCIT Model}
\label{fig:SCIT} 
\end{center}
\end{figure}

\par{{\it\textbf{Self-Cleansing Intrusion Tolerance (SCIT)}}:}
SCIT \cite{bangalore2009securing}, a method based on virtualization technology, tries to achieve mitigation by constantly cleansing the servers and rotating the roles of individual servers, as shown in Figure \ref{fig:SCIT}. We know that virtualization is a key technique in a cloud based environment. If a server is initiated, SCIT places a pristine, malware-free copy of the server's operating system into a virtual machine. Any server in the cluster switches between two modes periodically. The two modes are online servicing of clients and offline for cleansing. To coordinate among server modes, rotations can be performed with the help of a central controller or a distributed control mechanism using the Cluster Communication Protocol (CCP) \cite{huang2006scit}. In the rotation process, online servers are set offline. Afterwards, the system is rebooted to initiate cleansing procedures.

\par{{\it\textbf{Dynamic Resource Pricing}}:}
Mankins et al. \cite{mankins2001mitigating} discuss the applicability of dynamic resource pricing to discriminate good from bad traffic. They implement a dynamic pricing strategy that favors good user behavior and punishes aggressive adversarial behavior. They propose a distributed gateway framework and a payment protocol. The idea is to impose dynamically changing prices on both network servers and information resources so that the approach can push the cost of initiating service requests, in terms of monetary payments and/or computational burdens, to requesting clients. Thus, the architecture can provide for service quality discrimination to separate good client behavior from adversarial behavior in a private cloud environment serving a large set of heterogeneous consumers.

\par{{\it\textbf{Intelligent Fast-Flux Swarm Network}}:} Lua et al. \cite{lua2011mitigating} describe an intelligent fast-flux swarm network to mitigate DDoS attacks. This swarm network ensures autonomous coordination among nodes and allocation of swarm nodes (deploying nodes densely like bee colonies) to perform relay operations. They use the fast-flux hosting technique, which uses a very short Time-To-Live (TTL) parameter for any specific name record and reassigns host names at high frequency. A load-balancing process checks the health of nodes and removes those that are unresponsive. However, when a DDoS attack is in progress, it may not be robust. For better optimization, they use the intelligent water drop algorithm  \cite{shah2009intelligent}. The Intelligent Water Drop (IWD) algorithm is a nature inspired algorithm. The algorithm mimics how water drops behave in the flow of a river, i.e., the dynamic behavior of a river.

\par{{\it\textbf{Roaming Honeypot}}:}
Generally, honeypots are built in a network to trap malicious attackers. In traditional deployment, honeypots are situated in fixed locations and machines. However, having fixed locations makes the security of the entire operation vulnerable to sophisticated attacks. Khattab et al. \cite{khattab2004roaming} and  Sardana and Joshi \cite{sardana2009auto} propose the concept of roaming honeypots, changing the locations of the honeypots continuously and disguising them within a server pool. A subset of servers is active and provides service, while the rest of the server pool is idle and act as honeypots. The roaming honeypot scheme detects attacks from outside the firewall and mitigates attacks from behind the firewall by dropping all connections when a server switches from acting as honeypot to become an active server. So, if we can adapt this approach to the individual private cloud environment, a roaming honeypot may be a very good defender for that environment with limited resources for legitimate users.

\begin{figure*}[!h]
\captionsetup{font=normalsize}
\begin{center}
\includegraphics[scale=.55]{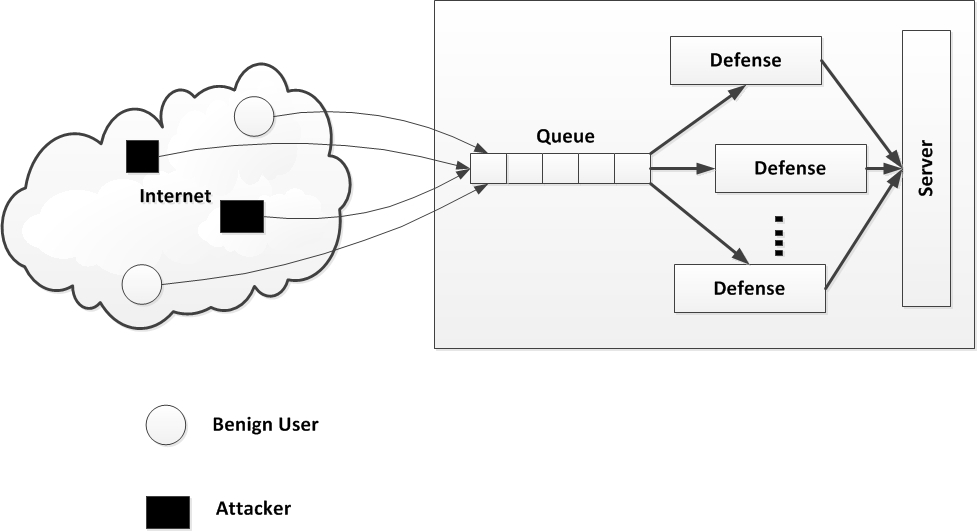}

\caption{Dynamic Resource Allocation Strategy by Yu et al. \cite{yu2014can}}
\label{fig:resourceallocation} 
\end{center}
\end{figure*}
\par{{\it\textbf{Moving Target Defense}}:}
Moving target defenses have been proposed as a way to make it much more difficult for an attacker to exploit a vulnerable system by changing aspects of that system to present attackers with a varying attack surface. The hope is that constructing a successful exploit requires analyzing properties of the system, and that in the time it takes an attacker to learn these properties and construct the exploit, the system will have changed enough so that by the time the attacker can launch the exploit to disrupt the exploit's functionality, the system has become more or less a new system \cite{touch2003dynabone, venkatesan2016moving}. This approach may provide an effective defense solution in context of private cloud environment as well.

\par{{\it\textbf{Dynamic Resource Allocation}}:}
In addition the traditional defense approaches, we need to explore resource allocation and utilization strategies for defending DDoS attacks in the cloud. Yau et al. \cite{yau2005defending} contend that DDoS defense is a resource management problem. Everyday the attack patterns keep changing. It will be a fruitless waste of time and resources to try to defend against DDoS attacks by just looking at patterns learned earlier. In addition, it is important to not only defend against an attack but also make services available during an attack. To beat DDoS attacks in the cloud, Yu et al. \cite{yu2014can} propose a dynamic resource allocation procedure within an individual cloud, as shown in Figure \ref{fig:resourceallocation}. It is a simple methodology of cloning Intrusion Prevention Servers (IPSs) from idle resources to filter out attack packets quickly and provide general services simultaneously. Some other specific resource allocation approaches have been proposed as well. We present them below.

\par 
Virtualization is a key concept in resource provisioning and management in the cloud. Virtualization provides a view of resources used to instantiate virtual machines. Isolating and migrating the state of a machine help improve optimization of resource allocation. Live virtual machine migration transfers the ``state" of a virtual machine from one physical machine to another, and can mitigate overload conditions and enable uninterrupted maintenance activities. Mishra et al. \cite{mishra2012dynamic} incorporate dynamic resource management in a virtual environment. Their approach answers basic questions such as when to migrate, how to migrate, types of migration and where to migrate. It also treats differently the migration of resources in different network architectures, e.g., LAN (Local Area Networks) and WAN (Wide Area Networks).
\par
The cloud environment can be described as probabilistic in nature. So there is a need to assess the performance of a cloud center for resource provisioning. The probabilistic nature of the cloud can be represented in terms of stochastic processes \cite{doob1953stochastic}. Shawky \cite{shawky2013performance} introduces an approach to model and analyze the performance of the resource allocation process using stochastic process algebra.

\begin{figure*}[!h]
\captionsetup{font=normalsize}
\begin{center}
\includegraphics[scale=.66]{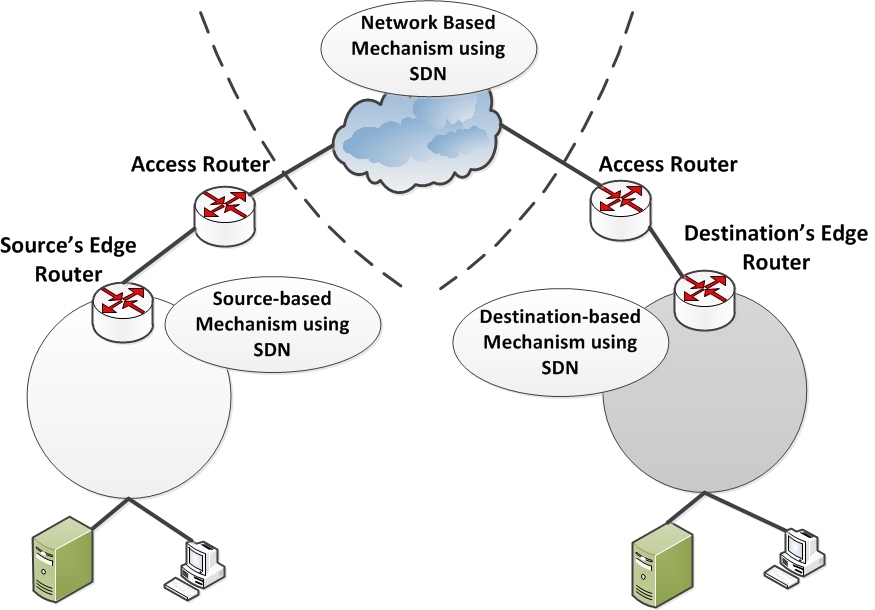}

\caption{Defense mechanisms against DDoS attack using SDN \cite{yan2016software}}
\label{fig:SDN} 
\end{center}
\end{figure*}

\par{{\it\textbf{SDN-based DDoS defense}}:}
The Software Defined Network (SDN) paradigm can be used to provide new opportunities to integrate application provisioning in the cloud with the network through programmable interfaces and automation \cite{banikazemi2013meridian}. The available options in SDNs (e.g., software-based traffic analysis, logical centralized control, global view of the network, and dynamic updating of forwarding rules) make it easy to provide detection and reaction to DDoS attacks in cloud environments. However, the separation of the control plane from the data plane in SDNs may introduce new attack planes. 
\par An SDN itself may be a target of some attacks, and potential DDoS vulnerabilities exist across SDN platforms. For example, an attacker can take advantage of the characteristics of SDNs to launch DDoS attacks against the control layer, infrastructure layer plane and application layer of SDNs. An attacker can infect a sufficient number of machines in a short time frame in traditional networks. On-demand self-service capabilities of the cloud that let legitimate businesses quickly add or subtract computing power could be used to instantly create a powerful botnet. Attackers are also known to use cloud as malware-as-a-service by renting different virtual machines and using them as bots. Separation of the control plane from the data plane enables one to establish easily large-scale attack and defense experiments. A logical centralized controller of an SDN permits a system defender to build consistent security policies and to monitor or analyze traffic patterns for potential security threats. A programmable intermediate network architecture can be setup easily in on an SDN. 
\par The cloud networks face challenges such as guaranteed performance of applications when applications are moved from on-premise to the cloud facility, flexible deployment of appliances (e.g., intrusion detection systems or firewalls), and security and privacy protection. An environment, providing good programmable, flexible and secure infrastructure is needed. SDNs are evolving as the key technology that can improve cloud manageability, scalability, controllability, and dynamism \cite{azodolmolky2013sdn}. In the past few years, several innovative SDN-based defense solutions have been introduced. These solutions belong to the three basic types of SDN-based DDoS defense mechanisms as shown in Figure \ref{fig:SDN}. In \cite{yan2016software} include a detailed discussion of SDNs, SDN-based clouds, and autonomous defense in clouds. SDNs can provide a new, dynamic network architecture that can transform traditional cloud network backbones into rich service-delivery platforms.

\par Lin et al. \cite{lin2014software} refer to SDNs as an emerging wave to transform network industries. They discuss SDNs and standardization in terms of components such as controllers, applications, service chains, network function virtualization and interfaces. SDN-based clouds are a new type cloud, in which SDN technology is used to establish control over network infrastructure and to provide networking-as-a-service (NaaS). In such clouds, cloud computing extends from server centralization and virtualization as well as storage centralization and virtualization to network centralization and virtualization. Banikazemi et al. \cite{banikazemi2013meridian} argue that service-level network models that provide higher-level connectivity and policy abstractions are integral parts of cloud applications. Yen and Su \cite{yen2014sdn} establish that an SDN-based cloud computing environment via open source OpenFLow switch and controller packages is effective in providing load balancing, power-saving and monitoring mechanisms. 

 A QoS-guaranteed approach is described in \cite{akella2014quality} for bandwidth allocation that satisfies QoS requirements for all priority cloud users by using Open vSwitch \cite{pfaff2013open} based on SDNs. An integrated solution is described in \cite{lin2013flow} to combine two strategies, flow migration and VM migration, to maximize throughput and minimize energy. Cziva et al. \cite{cziva2014sdn} present an SDN-based framework for live VM management where server hypervisors exploit temporal network information to migrate VMs and minimize the network-wide communication cost of the resulting traffic dynamics. In \cite{seeber2014improving}, authors claim that SDNs offer new opportunities for network security in cloud scenarios, because an SDN-based cloud provides more flexibility and faster reaction when the conditions are changing. Braga et al. \cite{braga2010lightweight} presents a lightweight method for DDoS attack detection based on traffic flow features, in which the extraction of such information is made with a very low overhead compared to traditional approaches. This is possible due to the use of the NOX platform  \cite{gude2008nox}, which provides a programmatic interface to facilitate the handling of switch information. Shin and Gu \cite{shin2013attacking}  show a new attack to fingerprint SDN networks and further launch efficient resource consumption attacks. This attack demonstrates that SDNs also introduce new security issues that may not be ignored. Flow Table Overloading in Software-Defined Networks is a vulnerablity to be handled carefully. Yuan et al. \cite{shendefending} points out this issue and provides a security service in an SDN using QoS-aware mitigation strategy, namely, peer support strategy, integrating the available idle flow table resource of the whole SDN system to mitigate such an attack on a single switch of the system.

 \par SDNs have been accepted as a new paradigm to provide an entire set of virtualization and control mechanisms to meet defense challenges in cloud networking. Thus, exploring the use of SDNs in providing better DDoS defense solutions in the cloud computing environment is likely to be beneficial.

\begin{figure}[h]
\captionsetup{font=normalsize}
\centering
\includegraphics[scale=.65]{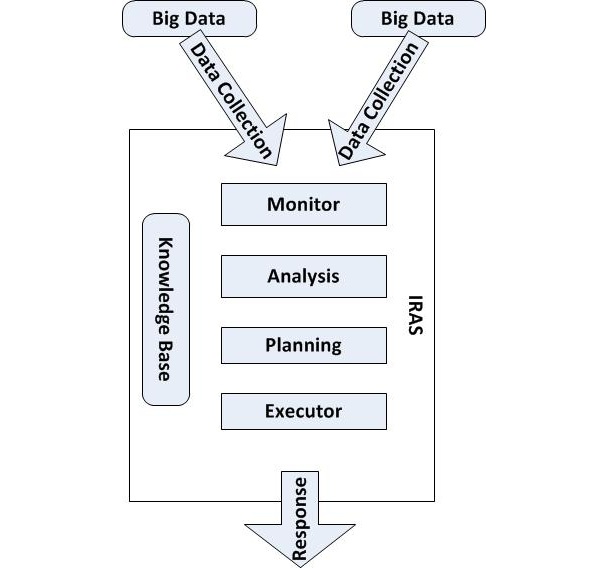}

\caption{Intrusion Responsive Autonomic System(IRAS)}
\label{fig:IRAS} 

\end{figure}

\begin{figure}[!h]
\captionsetup{font=normalsize}
\begin{center}
\includegraphics[scale=.575]{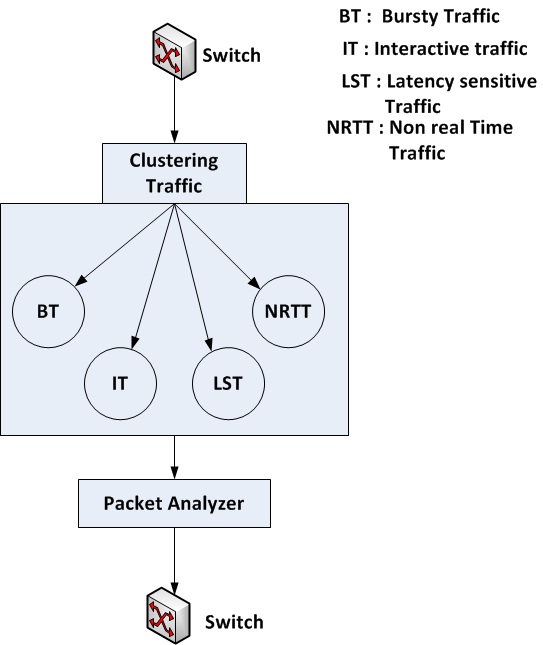}

\caption{Traffic Cluster Analysis by \cite{govinda2014secure}}
\label{fig:cluster} 
\end{center}
\end{figure}

\par{{\it\textbf{Big Data Analytics}:}
Anomaly detection is essential for preventing network outages and keeping the network resources available. However, to cope with the increasing growth of Internet traffic, network anomaly detectors are only exposed to sampled traffic, and as a result, harmful traffic may avoid detector examination. Fontugne et al. \cite{fontugne2014hashdoop} investigate the benefits of recent distributed computing approaches for real-time analysis of non-sampled Internet traffic. Their study is to detect network traffic anomalies using Hadoop. They also note that since MapReduce requires the dataset to be divided into small splits and anomaly detectors compute statistics from spatial and temporal traffic structures, special care should be taken when splitting traffic. They propose Hashdoop, a MapReduce framework that splits traffic with a hash function to preserve traffic structures.

\par Vieira et al. \cite{vieira2014autonomic} propose the Intrusion Responsive Autonomic System (IRAS) to analyze real time traffic to detect intrusion and mitigate attacks in the cloud platform, as shown in Figure \ref{fig:IRAS}. IRAS is an autonomous intrusion response technique endowed with self-awareness, self-optimization and self-healing properties. It runs through four steps, {\it monitor}, {\it analyze}, {\it plan} and {\it execute} to respond to the behavior patterns observed in real time big data using knowledge-based techniques. The sensors present in the system gather the log data from the network intrusion detection system and host systems.

\par As the Internet evolves and the computing infrastructure changes rapidly, the types of data being processed also evolve and change rapidly, and the complexity in structure and size of data being generated increases. All this is happening because more processing power produces more data at every opportunity. Researchers have coined the concept of ``Big Data" to refer to data handled by large enterprises like Google, Facebook, IBM and so on  \cite{lohr2012age}. Processing such data to gather information from a cloud network traffic is a big task. Big data traffic is collected, examined and analyzed in high performance servers to find interesting and useful patterns. The use of large scale distributed parallel processing of data in the cloud environment is commonplace. For example, Lee et al. \cite{lee2010internet} propose a method to analyze Internet traffic using the MapReduce  \cite{dean2008mapreduce} framework within the cloud computing platform. They compare their result with Hadoop \cite{white2009hadoop} and other tools concluding 72\% improvement in computational efficiency. Tripathi et al. \cite{tripathi2013hadoop} also study characteristics of DDoS attacks in the cloud and develope a scheme to detect such attacks in a Hadoop based environment. Lee et al. \cite{lee2011hadoop} also provide two algorithms to detect DDoS attacks using packet tracing method in a MapReduce environment.

\par It is obviously necessary to remove DDoS attack traffic from normal traffic in the cloud environment to reduce the burden of processing huge amounts of unwanted traffic, and to maximize the flow of normal traffic. Govinda and Sathiyamoorthy \cite{govinda2014secure} introduce a process of clustering the traffic into different groups. These groups are flash traffic, interactive traffic, latency sensitive traffic, non-real time traffic and unknown traffic, as shown in Figure \ref{fig:cluster}. They use Hadoop technology to analyze big data traffic. If any of these packets is categorized as unknown traffic, it is identified as a part of DDoS attack and eliminated by the packet analyzer.

\subsection{Discussion}
The approaches discussed in this section are presented compactly in Table \ref{tab:first}. We can summarize our discussions in the following observations. 
\begin{itemize}
\item It is necessary to build a real time defense system, whether it is network based or host based.
\item Incorporating dynamic behavior in the solution can provide adaptability to the defense.
\item The discussed methods employ the tolerance approach. Thus, allocating and utilizing resources effectively can provide a good defense.

\item As cloud computing systems incorporate traditional network topology and also new resource sharing methods, defense solutions against DDoS in the individual private cloud environment need to evolve to adapt to both. 
\item Resource utilization in a virtualized cloud computing environment is important. So, resource sharing and utilization need to be smooth enough to provide services along with security.
\item In a large infrastructure network, the converging network traffic will be always high enough for analysis. New data analysis techniques need to to be adapted for better defense.
\end{itemize}

\begin{table*}

\captionsetup{font=normalsize}
\caption{Selected Approaches Handling DDoS Attacks}
\label{tab:first}
\begin{center}
 \begin{tabular}{|p{3.5cm}  | p{7.5cm} | p{1cm} | p{2.04cm}|}

    \hline
    Authors   & Key Points & Real-Time & High rate/ Low rate \\
    \hline

Mankins et al. \cite{mankins2001mitigating}  &  Implements dynamic pricing strategy in terms of payments and/or computational load of each user. & Yes & Not mentioned \\

Ioannidis and Bellovin \cite{ioannidis2002implementing}  &  Uses a router based approach to detect and drop malicious packets. & Yes & High Rate\\
&  Achieves dropping of malicious packets by limiting the rate after matching attack signatures.&&\\

Kong et al. \cite{kong2003random}  &  Models the network flow to mitigate attacks. & Yes& Not\\
&  Expresses DDoS attacks in terms of max-flow min cut theorem \cite{dantzig2003max}.&& mentioned\\

Wang et al. \cite{wang2003sitar}  &  Provides minimal services to legitimate users in the critical moments during an attack.& Yes &Not mentioned\\
&  Use a dynamic fault tolerance architecture is to achieve the goal. &&\\

Khattab et al. \cite{khattab2004roaming}  &  Introduces a roaming honeypot technique to continuously disguise servers from the attacker, changing locations.& Yes& High rate\\
&  Provide general service using a subset of the servers and the rest of the idle servers act as honeypots. &&\\
&  Detects the attack and tries mitigation.  &&\\

Yauet al. \cite{yau2005defending} &  Throttles the traffic at upstream routers to forestall an impending attack. &Yes& High rate \\
&  Uses k-level max-min fairness theory. &&\\
&&&\\
Chen et al. \cite{chen2007divide} & Requests upstream routers of victim  to mark malicious packets for traceback. &Yes& Does not  \\
& Drops attack packets at the source end using traceback.  && depend on rate\\

Bangalore and Sood \cite{bangalore2009securing}&  Uses virtualization technology, cleansing and changing roles of servers to achieve mitigation. & Yes & Does not depend on rate\\
&  Reduces the server exposure time to network. && \\

Lua and Yow \cite{lua2011mitigating}&  Uses an intelligent fast-flux swarm network and adapts the intelligent water drop algorithm of \cite{shah2009intelligent}. & Yes & Does not depend on \\
&  Performs load balancing for optimization.&& rate \\

Lee et al. \cite{lee2011hadoop} &  Traces packet tracing in MapReduce environment. & Yes & Flow/Rate analysis\\

Yu et al. \cite{yu2014can}  &  Uses dynamic resource allocation policy. & Yes& Does not  \\

&  Perform reallocation or deallocation of resources for the intrusion prevention server based on the time required to compute and respond to each request packet.  && depend on rate\\

\hline
\end{tabular}
\end{center}
\end{table*}

\begin{table*}[!h]
\captionsetup{font=normalsize}

\begin{center}
 \begin{tabular}{|p{3.5cm}  | p{7.5cm} | p{1cm} | p{2.04cm}|}

    \hline
    Authors  & Key Points & Real-Time & High rate / Low rate \\
    \hline

Tripathi et al. \cite{tripathi2013hadoop}  &  Analyzes DDoS attack traffic patterns in the cloud environment.& Yes& Not mentioned\\
&  Perform detection of attacks in Hadoop environment. && \\

Vieira et al. \cite{vieira2014autonomic} &  Analyzes big data in real time to mitigate attacks in the cloud environment.& Yes& Not mentioned\\

Govinda and Sathiyamoorthy \cite{govinda2014secure} &  Provides a clustering technique on big data in the cloud environment to group different types of traffic.& Yes & Traffic analysis\\
&  Eliminates packets that are categorized as unknown traffic, by marking as DDoS attack.&&\\
\hline
\end{tabular}
\end{center}
\end{table*}
\subsection{Recommendations}
\label{sec:Recommendation}
In a private cloud environment, it is possible to build an effective defense solution against DDoS attacks. After analyzing many existing approaches, we can set some recommendations to adapt some of the discussed approaches and to develop the best feasible solutions. The cloud environment should have a dynamic firewall to detect abnormal changes in network traffic in real time. Like, SDN based solution can provide dynamic, cost-effective, adaptability and suitability for high bandwidth. If a preventive measure can work in cooperation with routers near the source router, the defense is likely to be stronger. We can explicitly reprogram all the cooperating routers to create centralized or distributed defense using SDN paradigm according to our need. This way we may be able to trace back the source of the attack or provide a defense as near as possible towards the source-end. SDN controller applications are mostly deployed in large-scale scenarios. A huge attack of size in terabit per second need to defend in cloud environment. A distributed and cooperative agent-based DDoS tolerant architecture can help counter that kind of huge DDoS attacks in real time.

\section{Challenges and Issues}
\label{sec:Challenges}
 A service provider usually has an adequate amount resources for specific service seekers. The sophistication of cloud architectures and virtualized abstraction of the resources have introduced issues and challenges in deploying effective design solutions to defend against DDoS attacks suitable for individual private cloud platforms. In a private cloud, customers share computing and store resources paying on a per use or subscription basis, and so frequent DDoS attacks can jeopardize the entire operation. There are a few challenges and issues to be addressed to make effective use of the cloud when defending against DDoS attacks. 

\begin{itemize}
\item For a cyber defense tool, effectiveness should be measured in terms of time taken and accuracy of detection obtained in real time. If not addressed properly before deployment and thoroughly tested, lack of efficient performance can be a roadblock to large-scale adoption of any real-time defense mechanism. This is especially true in our case because the architecture of the cloud demands a different DDoS defense model rather than a traditional one in a general network.

\item Vulnerability or flaws in protocol or policy in the cloud environment is another major reason for flooding DDoS attacks. A mitigation technique for flooding attacks must take into account system and protocol design to ensure an effective and successful implementation.

\item All cloud users share the same pool of resources. This makes it absolutely essential to start with requirements that ensure reliability, security and separation issues from the outset. This has not been well considered in traditional DDoS attack defense. The service provider must ensure that its DDoS attack defense operations neither affect nor are affected by other cloud activities.

\item If the cloud provider has only the resources required to provide services to its customers but not much more to defend, this may encourage undesirable DDoS attacks if attackers can guess the situation. This way, the system could give out inflated statistics to the outside world.

\item In a private cloud environment, defending against a DDoS attack is more about resource management. The challenge is to be able to perform rapid re-allocation or use a dynamic network topology. It may make the attack traffic more difficult to handle because the defense mechanism may need to update the network by changing physical locations of virtual machines. We need to build the defense strategy keeping in mind that live migration technology  \cite{clark2005live} enables faster execution of the needs imposed by the strategy.

\item Skillful resource allocation and virtual machine migration lead to frequent topological changes in the network from the defender's viewpoint. Such  resource allocation and virtual machine migration processes are fast-paced. Thus, an approach to defense against DDoS attacks must be able to adapt to a dynamic network with frequent topological changes and still maintain high detection rate and prompt reaction capability. In other words, a successful defense mechanism must be dynamic and adaptive.

\item It is necessary to detect an attack quickly with minimal required information to reduce communication overhead. Quick detection of an attack ensures that the next phase of mitigating the attack can avail itself of the time and resources needed in a dynamic framework. The detection algorithm needs to match its speed with the packet forwarding scheme used in the cloud technology. Using a traditional detection mechanism like signature-based or anomaly-based approaches or using both, the issue may be addressed.

\begin{figure*}[]
\captionsetup{font=normalsize}
\begin{center}
\includegraphics[scale=.575]{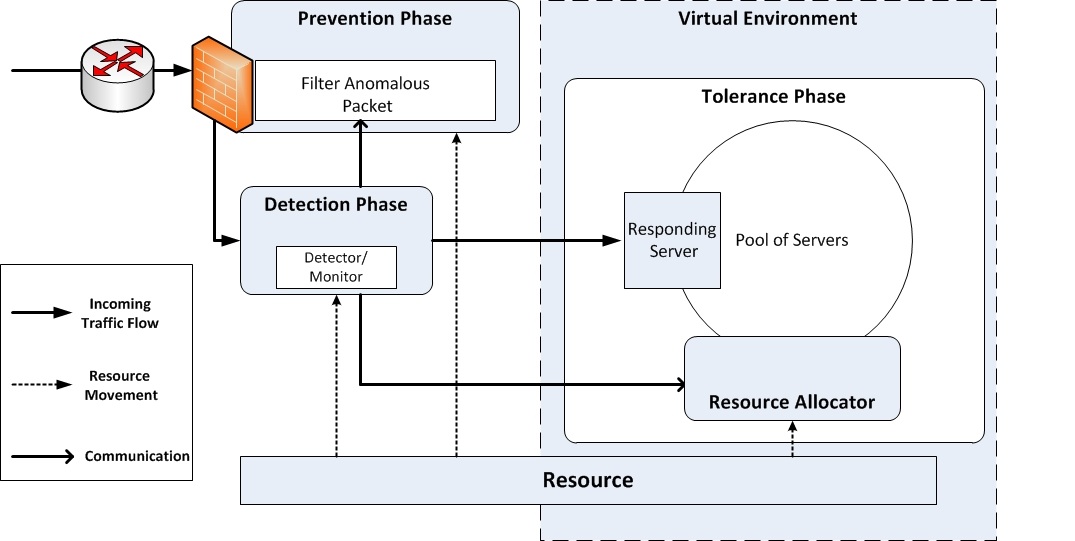}

\caption{A Generic Cloud Based Defense Framework}
\label{fig:framework} 
\end{center}
\end{figure*}

\item There are no common characteristics among traffic
streams comprising various attacks. Patterns for different attacks are different. It is obvious that one cannot build defensive approaches for each type of attack in a private cloud with a particular amount of resource dedicated to each attack. Thus, it is important to build a generic architecture to defend against most types of DDoS attacks.

\item No security precautions can guarantee that a system will never be intruded and so at the critical moment when the system is designed, applications still need to provide minimal services to the legitimate users even under active attacks or when partially compromised.
\end{itemize}

\section{A Generic Framework}

\label{sec:framework}
  Based on the recommendations presented earlier, we believe that an automatic host based approach emphasizing tolerance can provide better utilization of resources in the cloud environment to respond to DDoS attacks in an individual private cloud. With limited resources, it is necessary to develop a procedure to defend against DDoS attacks and to provide general service. A generic conceptual framework is shown in Figure \ref{fig:framework}. It is a combination of different phases and components. The whole defense module is just a conceptual depiction of cloud based defense solution against DDoS attacks adapting concepts borrowed from existing techniques, adapted to a new environment. Detection and prevention phases of this framework incorporate some ideas of the traditional Internet and also tolerance techniques to the cloud environment. The framework, that we discuss below, abides by the recommendations discussed previously.

\begin{itemize}
\item[(i)] Detection Phase: In this phase, the {\it monitor} component analyzes the behavior of the traffic coming to a responding server which handles incoming requests. If the incoming traffic shows any abnormality, the monitor catches it automatically and sends an alert message to the {\it resource allocator}. The alert message contains the threat level, how to act to tackle the abnormality and when to initiate the migration stage. The monitor will also communicate with the prevention component with alert messages about the incoming traffic. This component needs to detect abnormal changes in network traffic in real time.
\item[(ii)] Tolerance Phase: In this phase, we can utilize the resources effectively using an virtualizaton technique available in the cloud environment. For example, the data center or the private cloud provider has the ability to provide the resources to users using virtualization. So, the utilization of the resources should be appropriate for the security needed in a crisis situation since resources are always limited in private cloud environment.
\begin{itemize}
\item The \textit{resource allocator} maintains a queue of fresh server copies to provide services that must be rendered by the responding server. Depending on the level of the alert message, it tries to maintain a dynamic queue of spare resources to fight back if a rapid change in service demands occur because of any high traffic attack. It can also push unnecessary resources back to the resource pool when the state becomes normal.
\item A responding server may be detached from service depending on two things, time and computation load. A server needs to be exposed only for a limited amount of time and if the computation load exceeds a threshold level due to malicious activity, it can be switched to inactive status. Before switching, using live migration we can copy the necessary states of the server to an incoming fresh server so that the usual services can be resumed with minimum delay.
\end{itemize}
\item[(iii)] Prevention Phase: If a preventive measure can work in cooperation with routers near the source router, the defense is likely to be stronger. An adaptive and dynamic mapping intrusion response system for effective prevention of DDoS attacks in real time is essential. In the prevention phase, alert messages coming from the detection component need to be analyzed and correlated to discover patterns or strategy in attacks. Using these, we can filter out matching incoming packets later with a low false positive rate. If the traffic flow is high, the amount of alert messages may be high enough to analyze. In such a situation, we need to use newly developed data analysis techniques, such as big data analytics to analyze the patterns.   
\end{itemize}
\section{Conclusion and Future Work}
\label{sec:Conclusion}
We can definitely say that in the near future, most computing activities and resources will migrate to the cloud and security will be a prime concern. DDoS attacks may be resisted with generic solutions to survive and to provide best services under the circumstances. However, to be successful, more than the usual is necessary in the cloud context. In this paper, we have discussed issues in handling DDoS attacks, specifically in a private cloud environment. We have highlighted issues and challenges faced in the private cloud environment when providing defense solutions against DDoS attacks. Some useful approaches developed by researchers to address these issues have been presented and analyzed in this paper. The importance of mitigating the attack by tolerating it and by optimized use of resources in the private cloud scenario has been emphasized. Finally, the role of big data analytics in defending DDoS attacks in the cloud has been introduced.
\par	In the near future, we plan to deploy the conceptual cloud framework in a testbed to demonstrate and analyze the effectiveness of our proposed solutions. It is important to know how far this framework can resolve different issues and challenges when defending against DDoS attacks in an individual private cloud environment.

\bibliographystyle{wileyj}


\end{document}